\documentclass[twocolumn,showpacs,amsmath, amssymb,prb]{revtex4}
\usepackage{graphicx}
\begin{document}
\title{Commensurability Transition and Stripe Phases in the Ginzburg-Landau Theory}
\author{B. Uchoa and G. G. Cabrera}
\address{Instituto de F\'{\i }sica `Gleb Wataghin',\\
Universidade Estadual de Campinas (UNICAMP),\\
C. P. 6165, Campinas 13083-970 SP, Brazil}
\date{\today}

\begin{abstract}
We phenomenologically describe the thermodynamics of charge and spin density
waves in doped high-$T_{c}$ oxides. We have explicitly calculated stable
non-homogeneous solutions in the incompressible spin driven stripe phase,
where stripes are static soliton-like charge density waves (CDW), and in the
vicinity of their critical point, where CDW's become harmonic. Our phase
diagram points to a commensurability transition separating the low (LI) and
high (HI) incommensurable phases. Besides, we demonstrate by rigorous group
symmetry arguments that the stripe criticality is compatible with a second
order phase transition.
\end{abstract}

\maketitle

\narrowtext

\section{Introduction}

Stripe formation in the high-$T_{c}$ oxides has been widely considered by
many theorists in the recent years. The present experimental evidences point
to a curious electronic order, where carriers introduced by doping (holes)
segregate into `rivers of charge'. Those structures, generically called {\em %
stripes}, are considered as `nearly one dimensional' objects immersed in an
insulating antiferromagnetic sea of spins. This picture was inferred more
accurately after recent neutron diffraction experiments \cite{trb,tr96},
which observed incommensurable antiferromagnetic domains separated by charge
stripes localized in the N\'{e}el wall regions. It is found that the
magnetic scattering peaks define a superzone which is half the size of the
one associated with the charge order, indicating that stripes introduce a $%
\pi $-phase shift between neighboring magnetic domains. The above structure
results from competing effects, which in conjunction, reduce the energy cost
for breaking antiferromagnetic bonds and minimize the kinetic energy of
holes.

The incommensurability of the peaks with the reciprocal lattice is a strong
argument in favor of the idea that holes condense in stripes of finite width
oriented along a given direction in the insulating planes. The temperature
dependence of the peak splitting, showing that in some systems the
charge-ordered peaks appear at higher temperatures than the magnetic ones 
\cite{trb}, also discards the hypothesis of nested Fermi surfaces.
Currently, the stripe picture is considered as a good candidate for
explaining many of the properties (normal and superconducting) of the high-$%
T_{c}$ and other layered oxides\cite{stripes}. In contrast, there is still a
lot of controversy about the nature of the correct spin order at the
N\'{e}el walls, {\it i.e} spiral or collinear. Both of them are possible
under the available experimental data \cite{trb,tra}, despite theoretical
difficulties to justify spiral order for the ground state \cite{pr}.

Another key point concerning the stability of stripes is their robustness in
the absence of long-range antiferromagnetism, at temperatures considerably
above of the three-dimensional N\'{e}el ordering. They even survive the
absence of two-dimensional antiferromagnetic correlations, which occurs at
higher temperature, showing that the transition is charge-driven\cite{length}%
. This phenomenon is ascribed to the separation of charge and spin
fluctuations \cite{zaa3}, which establishes strong dynamic correlations that
lower the large energy due to the hole repulsion inside the stripes. This
behavior is quite ubiquitous in all oxide families, except for the chromium
oxides, where the stripe and spin orders collapse together \cite{py}.

The current idea among theorists (originally suggested by Anderson \cite
{Anderson} ) is that a sufficiently large repulsive on-site $U$ potential
might allow for antiferromagnetism in the oxygen planes. The antiparallel
alignment of the antiferromagnetic ordering at half filling, lowers
perturbatively the kinetic energy in doped samples, because carriers (holes
in most cases) can hop to neighboring sites. Antiferromagnetism also produce
an attractive force among holes, in order to minimize the number of broken
antiferromagnetic bonds. The competition between this short range attraction
and the long range Coulomb repulsion of holes, may stabilize the
non-homogeneous phase of stripes, frustrating the tendency to macroscopic
phase separation. From the view point of continuous models, the hole
repulsion competes with the ``helicity'' of the spiral spin order , meaning
that the antiphase of domains is crucial to achieve the thermodynamic
stability of the stripe regime\cite{pr}. The other possibility, magnetic
domains in phase, should be unstable in relation to the homogeneous state.
This fact, also rules out the Ising limit, with very narrow walls between
domains.

The physics involved in the present problem, embraces as much the long range
scale of charge interactions as the microscopic scale of spin fluctuations,
where theoretical considerations in the basis of a macroscopic mean field
may seem meaningless at first sight. However, for the stripe regime,
spin-spin and charge-charge correlation lengths should be comparable in a
certain region of doping and temperature, with non-vanishing two-dimensional
antiferromagnetic correlations. In spite that long range antiferromagnetic
order is not attained in two dimensions, typical correlation lengths are of
the order of $100-200$ lattice parameters\cite{length,Lee}. We then define a
two-dimensional $T_{N}$ as the temperature where the magnetic neutron
diffraction peaks are not resolved any more\cite{RMP}. In a similar way, a
critical point $T_{S}$ for the stripe order is assumed\cite{tr96}. The above
facts justify a theoretical treatment using the Ginzburg-Landau (GL)
approach to elucidate the internal structure of stripes.

\begin{figure}[h!]
{\centering \resizebox*{2.4 in}{!}{\includegraphics{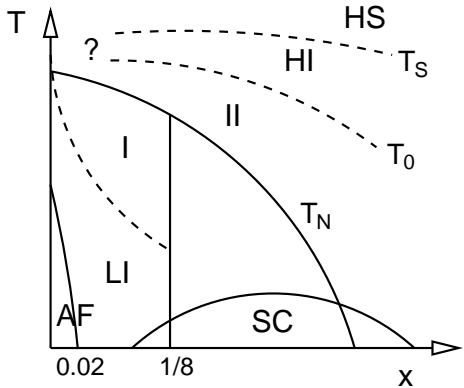}} \par}
\caption{\label{fig1} Cuprate phase diagram temperature vs doping. AF is the true
long range antiferromagnetic order, below $x\sim 0.02$. Inside the quarter
of dome is the 2D (quasi-long range) antiferromagnetic phase,
coexisting with the superconducting one (SC). LI is the spin driven Low
Incommensurable phase of stripes. Regions I and II are High Incommensurable
phases with and without spin order, respectively. The harmonic High
Incommensurable phase is denoted by HI, right below the High Symmetry (HS)
phase, where stripes disappear. $T_0$ indicates a second order 
transition separating high and low incommensurable phases.}
\end{figure}

The central idea of this paper is to assess the properties of the phase
diagram, shown in a qualitative fashion in Fig. \ref{fig1}, and to theorize about the
nature of the corresponding phase transitions. We will concentrate our study
in two regions, where controversial statements are found in the literature,
namely region LI, for the incompressible stripe phase below $T_{N}$, and
region HI, for the paramagnetic phase immediately below the stripe critical
point $T_{S}$. We will show that the charge distribution suffers a
transition of commensurability, which evolves from the low incommensurable
regime in phase LI, characterized by a typical soliton-like distribution, to
a highly incommensurable one, which extends until phase HI and collapses at $%
T_{S}$, following possibly a second order transition.

In the first step of
the calculation, we study the marginal stripe phase transition below $T_{N}$%
, building a GL free energy with the staggered magnetization as the main
order parameter (OP) and the charge density playing the role of a secondary
OP. This construction is based on a microscopic picture, which is
supplemented by symmetry considerations. In contrast, in the second part of
the paper, we study the highly incommensurable phase in the absence of
antiferromagnetism, the transition being driven by the charge density. We
develop theoretical arguments entirely based on symmetry to build the GL
free energy for the HI phase, where the CDW is harmonic in the vicinity of $T_S$. 
We calculate the
amplitude and wavelength of the CDW and estimate the limits of the harmonic
approximation. This one and other conjectured phases are displayed in the
diagram of Fig.1. We end the paper with a discussion of our results.

\section{Low incommensurable stripe phase}

Attention has been paid by some theorists to the tendency of attraction
between domain walls at large distance. Priadko and coworkers \cite{pr}
believe that this fact leads to an ubiquitous tendency to phase segregation
that renders the stripes unstable in the small doping limit. They
demonstrated the existence of an attractive force between asymptotically
separated charged domain walls when the charge effect is considered in first
order by means of a chemical potential. Here, we argue that the spin
exchange ``traps'' the holes inside domain walls, leading to a direct
competition between the spin order attraction and the long range repulsion
of holes, which stabilizes the inter-stripe separation. In spite that
charges play the role of a secondary OP, their interactions must be included
at least in second order in the free energy. In the cuprate case, it is
experimentally known that for the doping range $0.05<x<\frac{1}{8}$, the
effect of doping increasing is counterbalanced by the reduction of the
domain width $d$, and {\em vice-versa}, following the empirical relation $%
x\sim \frac{1}{d}$. In this phase, the stripes are internally charge
incompressible and the low doping regime may be simply thought as a
re-scaling of the non-interacting stripe system.

First of all, we propose a constraint in the form of a global charge
conservation 
\[
\int \!{\rm D}r\,\rho ({\bf r})=Q\ , 
\]
where the integration is performed over a two dimensional system with linear
dimensions $(l,l_{z})$, with the $z$-axis along the charge modulation
direction. From now on, the charge of holes, the electron spin, and the
lattice constant are normalized to unity and will be omitted. If the stripes
are non-interacting, the system can be rescaled along the direction $z$ in
terms of the domain width $d$, considering that two neighboring stripes are
infinitely distant ({\it i.e. }$d\rightarrow \infty $). This way, the charge
conservation principle can be simply restated as

\begin{equation}
\Pi \equiv \lim_{d\to \infty }\int_{-d/2}^{d/2}\!\!{\rm d}z\,\,\rho
(z)=h(T)\ ,  \label{constraint1}
\end{equation}
where $h(T)$ is a scalar function of temperature, $h(T)$ being proportional
to a linear charge density $Q/l$ divided by the number of domains $l_{z}/d$
along the modulation direction, {\it i.e}. 
\[
h(T)=\frac{Q}{ll_{z}}d=x\,d\quad , 
\]
or $x\propto \frac{1}{d}$. Therefore, the constraint (\ref{constraint1}) is
quite appropriate for the low doping regime.

The total energy of each plane must take into account the long range
potential for holes along with the exchange interaction of spins. The
simplest static Hamiltonian for a square lattice of N sites is 
\begin{eqnarray}
H &=&H_{c}+H_{s}=\,\sum_{i,\delta }^{N}\,\frac{1}{\epsilon _{\delta }}%
\,(1-n_{i})(1-n_{i+\delta })\,+  \nonumber \\
&&\qquad \qquad -\,\sum_{i}^{N}\,J_{i}\,n_{i}(\alpha _{1}S_{1\,i}^{2}+\alpha
_{2}S_{2\,i}^{2})\,+  \nonumber \\
&&\sum_{i,\delta }^{N}\,J_{i,\delta }\,n_{i}n_{i+\delta }(\beta
_{1}S_{1\,i}S_{1\,i+\delta }+\beta _{2}S_{2\,i}S_{2\,i+\delta })\quad ,
\label{hamiltonian1}
\end{eqnarray}
with $n_{i}=0,1,$ the spin occupation number. The interactions are extended
to distant neighbors, which are labeled by $\delta $. The first term in (\ref
{hamiltonian1}) yields the repulsive interactions of (spinless) holes with
coupling $\left( 1/\epsilon _{\delta }\right) $. The second term includes
the crystal field contribution, with anisotropy parameters $\alpha _{1}$ and 
$\alpha _{2}.$ The last term considers the exchange interactions among spins
($J_{i,\delta }>0$), which in general are assumed to be anisotropic, with
parameters $\beta _{1}$ and $\beta _{2}$. The spins are assumed to be
classical variables and are constrained to lay on the plane. Measurements
taken in a considerable variety of layered compounds reveal that the inplane
spin exchange interactions are isotropic to a high accuracy, meaning that
the magnetic crystalline anisotropy dominates\cite{aniso}. Therefore, we
consider $\beta _{1}\sim \beta _{2}$ and adopt $\alpha _{1}$ and $\alpha
_{2} $ as the source of anisotropy in our model.

To take the continuous limit of Eq. (\ref{hamiltonian1}), we partition
the antiferromagnetic lattice in two ferromagnetic sublattices, and define a
mean field for each sublattice as ${\bf m}_{\uparrow \downarrow }({\bf r}%
)=\left\langle n_{i}{\bf S}_{i}\right\rangle _{B({\bf r})},$ calculating the
spin average in a ball centered at the point ${\bf r}$, in real space. In
terms of the staggered fields $\left( {\bf m}_{\uparrow },{\bf m}%
_{\downarrow }\right) $, we define a continuous spin OP as 
\[
{\bf m}({\bf r})\equiv \left( {\bf m}_{\uparrow }-{\bf m}_{\downarrow
}\right) ({\bf r})\ , 
\]
in the limit when the lattice constant goes to zero. Its norm $m$ is a
positive number between $0$ and $1$. In the absence of dynamics, ${\bf \ m}%
_{\uparrow }$ and ${\bf \ m}_{\downarrow }$ are locally antiparallel and thus 
$m$ is maximum in the saturation regime, for $T\ll T_{N}$. We also define a
renormalized spin occupation density $\eta $, varying in the interval $[0,1]$%
, which plays the role of the secondary OP (it does not vanish at $T_{N}$%
). To remove the criticality at $T_{N}$, we impose 
\begin{equation}
\eta ({\bf r})=m({\bf r})\,t^{-\beta }\quad ,  \label{constraint2}
\end{equation}
where $t\equiv 1-T/T_{N}$ is the reduced temperature, and $\beta >0$
corresponds to the OP critical exponent. If we consider now the OP
components along the directions parallel and transverse to the stripes,
meaning $m_{\Vert }=m\cos \theta $ and $m_{\bot }=m\sin \theta $, and use
relation (\ref{constraint2}), we find that the macroscopic regime is well
defined in terms of the two OP degrees of freedom, $\eta $ and $\theta $.
The symmetry group of the OP is the subgroup $C_{2v}$ of $O(2)$, which
includes inversion of the spins and two reflection planes along the two
principal axes of the magnetization. The continuous limit of the microscopic
Hamiltonian (\ref{hamiltonian1}) leads to the GL free energy density (see
the Appendix) 
\begin{eqnarray}
f({\bf r}) &=&\,\,(t^{2\beta }b_{o}-a_{o})(\dot{\eta})^{2}\,+\,t^{2\beta
}b_{o}(\dot{\theta})^{2}\,\eta ^{2}\,-\,2a_{2}\eta  \nonumber \\
&+&\left[ a_{2}+c_{2}t^{2\beta }-\,t^{2\beta }b_{2}\left( \alpha _{\Vert
}\cos ^{2}\theta +\alpha _{\bot }\sin ^{2}\theta \right) \right] \eta ^{2} 
\nonumber \\
&&\qquad -\,2c_{2}t^{2\beta }\eta ^{3}+\left[ c_{2}\,t^{2\beta }+t^{4\beta
}b_{4}\right] \eta ^{4}\quad .  \label{freeEnergy2}
\end{eqnarray}
The dot means a spatial derivative with respect to $z$. We assume $a_{2}$
and $c_{2}$ to be regular functions of temperature, say $a_{2}(T)$ and $%
c_{2}(T)$, while $b_{2}\equiv bt$ is a standard Landau second order term
parameter, with $b$ a positive constant.

To satisfy the continuity condition of the magnetic field in the stripe
interface with the antiferromagnetic environment, we must have ${\bf m}\cdot 
{\bf \hat{e}}_{\bot }=0$ inside the bulk of domains. It leads to fixing the
direction of easiest magnetization along ${\bf \hat{e}}_{\Vert }$, what is
just equivalent to impose $\alpha _{\bot }<\alpha _{\Vert }$ to the
anisotropy parameters. The $C_{2v}$ point group of the spin ordered phase
admits two invariant terms, $m^{2}$ and $m^{2}\!\cos (2\theta )$, which
generate the polynomial expansion of the free energy potential, \cite{to}.
The second order anisotropy term $b_{2}m^{2}(\alpha _{\bot }\sin ^{2}\theta
+\alpha _{\Vert }\cos ^{2}\theta )$ can be decomposed in such basis as $%
\frac{1}{2}b_{2}\left\{ (\alpha _{\Vert }+\alpha _{\bot })m^{2}+(\alpha
_{\Vert }-\alpha _{\bot })m^{2}\!\cos (2\theta )\right\} $, making evident
the competition between the full symmetry term of the paramagnetic phase ($%
O(2)$ point group) and the $C_{2v}$ symmetry one. From this, the proper
expression for the inplane crystalline anisotropy is 
\begin{equation}
0<\varepsilon \equiv \frac{\alpha _{\Vert }-\alpha _{\bot }}{\alpha _{\Vert
}+\alpha _{\bot }}\leq 1\quad .
\end{equation}

The appropriate set of boundary conditions (BC) that applies to the free
energy (\ref{freeEnergy2}) is 
\begin{eqnarray}
\theta \! &=&\!\left\{ 
\begin{array}{ll}
0 & z\rightarrow -\infty \\ 
\pi & z\rightarrow \infty
\end{array}
\right. \\
&&  \nonumber \\
\dot{\theta}\! &=&\!0,\qquad z\rightarrow \pm \infty \,
\end{eqnarray}
and 
\begin{eqnarray}
\eta \! &=&\!1,\quad \,z\rightarrow \pm \infty \qquad  \label{BC3} \\
\dot{\eta}\! &=&\!0,\quad \,z\rightarrow \pm \infty  \label{BC4}
\end{eqnarray}

The SDW modulates transversely to the stripes, in order to satisfy $\rho
=1-\eta $. If we denote the energy of a single domain by $E\equiv
\int_{-\infty }^{\infty }f\,{\rm d}z$ , variation under the constrain of the
global charge conservation (\ref{constraint1}), yields 
\[
\delta (E+\Lambda \,\Pi )=\delta \int \!{\rm d}z\left( f(\theta ,\eta ,\dot{%
\theta},\dot{\eta},z)-\Lambda \,\eta \right) =0\quad ,
\]
where $\Lambda $ is a Lagrange multiplier. The variational equation above
results in 
\begin{equation}
\cos \theta =-\tanh \left( \lambda ^{-1}z\right) \quad ,
\label{TetaSolution}
\end{equation}
and 
\begin{eqnarray}
A\,\ddot{\eta} &-&\,\left( S-U\,{\rm sech}^{2}(\lambda ^{-1}\,z)\right) \eta
-3C(1-\,\eta ^{2})\,+  \nonumber \\
&&\qquad +2D\,(1-\eta ^{3})\,+\,S=0\quad ,  \label{ELeta}
\end{eqnarray}
where $\lambda ^{-1}=[(\alpha _{\Vert }-\alpha _{\bot })b_{2}/b_{0}]^{\frac{1%
}{2}}$ is the domain wall width in the spiral order. Similarly to Bloch
walls in ferromagnets, a finite $\lambda $ results from the competition
between the exchange and the anisotropy terms in the magnetic energy \cite
{condmat}. In expression (\ref{ELeta}), we have the quantities
\begin{eqnarray*}
A &=&b_{0}t^{2\beta }-a_{0}\ , \\
S &=&a_{2}-t^{2\beta }b_{2}\alpha _{\Vert }+c_{2}\,t^{2\beta }\ , \\
U &=&-2(\alpha _{\Vert }-\alpha _{\bot })\,b_{2}\,t^{2\beta }\ , \\
C &=&c_{2}\,t^{2\beta }\ , \\
D &=&c_{2}\,t^{2\beta }+t^{4\beta }b_{4}\ ,
\end{eqnarray*}
which are defined in terms of the original GL parameters given in (\ref
{freeEnergy2}). As usual, $a_{0}$ and $b_{0}$ have no dependence on
temperature and are assumed to be independent of $(a_{2},\,b_{2},\,c_{2}t^{2%
\beta },b_{4}t^{4\beta })$ in the parameter space. The limit $%
a_{0},b_{0}\rightarrow 0$, with $\lambda ^{-1}\rightarrow \infty $,
reproduces the Ising case where (\ref{ELeta}) becomes a $z$-independent
equation 
\begin{equation}
S(1-\eta )-3C(1-\,\eta ^{2})\,+\,2D\,(1-\eta ^{3})=0\,,  \label{z}
\end{equation}
which yields a constant $\eta =1-x$. Substituting $\eta $ in terms of the
doping $x\,$ in (\ref{z}) one finds 
\begin{equation}
S=3C\,(2-x)-2D\,\left[ 3(1-x)+x^{2}\right] \,\quad ,  \label{condition1}
\end{equation}
for $x\neq 0$. In order (\ref{condition1}) to be valid in a continuous
interval of temperature, we must have $a_{2}(T)\propto t^{2}$, $%
c_{2}(T)\propto t$ and the OP critical exponent $\beta =\frac{1}{2}$.

Now we solve equation (\ref{ELeta}) in a perturbative way, in relation to
the parameters $(c_{2},b_{4})$ which are associated to the cubic and quartic
terms in the free energy density (\ref{freeEnergy2}). We write 
\[
\eta (z)=\eta _{L}(z)+c\,\psi (z), 
\]
with $\eta _{L}$ satisfying eq. (\ref{ELeta}) in lowest order and $\psi $
corresponding to higher order corrections due to $c_{2}$ and $b_{4}$, which
are small parameters in the GL sense. Imposing the limit $c_{2}$, $%
b_{4}\rightarrow 0$, condition (\ref{condition1}) becomes $S=0$ for $x\neq 0$%
, and we obtain from (\ref{ELeta}) 
\begin{equation}
A\,\ddot{\eta}_{L}+U\,{\rm sech}^{2}(\lambda ^{-1}\,z)\eta _{L}=0\quad ,
\label{ELeta2}
\end{equation}
whose solution is the hypergeometric function 
\begin{equation}
\eta _{L}(z)=F\left( \,\Gamma _{-}\,,\Gamma _{+}\,;\,1\,;\,{\rm sech}%
^{2}(\lambda ^{-1}\,z)\,\right) \quad ,  \label{EtaSolution}
\end{equation}
with $\Gamma _{\pm }=\frac{1}{4}(1\pm \sqrt{1+4\,U\lambda ^{2}/A})$. Solving
the linearized $\psi $ equation resulting from (\ref{ELeta}) in the
asymptotic limit $\eta _{L}\sim 1$, where $\eta _{L}$ satisfies the equation
(\ref{ELeta2}), we encounter that 
\begin{equation}
\psi (z)\sim g(z)\left[ {\rm sech}(\lambda ^{-1}z)\right] ^{\lambda \sqrt{V}%
}\quad ,
\end{equation}
where $g(z)$ is a smooth and convergent hypergeometric function 
\[
g(z)=F\left( \,\Lambda _{-}\,,\Lambda _{+}\,;\,1+\lambda \sqrt{V}\,;\,{\rm %
sech}^{2}(\lambda ^{-1}\,z)\,\right) \quad , 
\]
with $\Lambda _{\pm }=\Gamma _{\pm }+\frac{\lambda }{2}\sqrt{V}$ and 
\[
V=\frac{1}{A}\,x\left[ 2D(3-x)-3C\right] 
\]
as an explicitly doping dependent quantity.

Going one more step further, we apply the constraint (\ref{constraint1}) to
the perturbed solution $\eta _{L}+c\,\psi $, which yields 
\begin{equation}
c=\frac{\int_{-\infty }^{\infty }\!{\rm d}u\,\left[ 1-\eta _{L}(u)\right] -x%
\frac{d}{\lambda }}{\int_{-\infty }^{\infty }\!{\rm d}u\,\psi (u)}\quad ,
\end{equation}
where $u=z/\lambda $ is the coordinate scaled by the soliton width.

\subsection{Estimation of parameters}

A numerical calculation shows that $\int_{-\infty }^{\infty }\!{\rm d}%
u\,(1-\eta _{L})\cong 0.81\,U\lambda ^{2}/A$, with $u=z/\lambda $. In lowest
order we write the constraint (\ref{constraint1}) as $\int \!{\rm d}%
u\,(1-\eta _{L})=\frac{d}{\lambda }x$, and find that 
\begin{equation}
\frac{d}{\lambda }x\sim \frac{4}{5}\frac{U}{A}\lambda ^{2}=\frac{8}{5}\frac{%
b_{0}t}{a_{0}-b_{0}t}\quad .  \label{inequality1}
\end{equation}
Up to this order, the amplitude of the CDW peaks depends on the adimensional
parameter $U\lambda ^{2}/A$ only. This latter quantity is numerically
limited to the interval $[0,2]$ in order to have $0\leq \rho (z)\leq 1$. As
a consequence, relation (\ref{inequality1}) leads to the approximate
inequality 
\[
0\leq \frac{d}{\lambda }x\lesssim \frac{8}{5}. 
\]
The above condition also allows us to estimate the limitation of the
temperature to saturate the peak. Within our simplified model we get $%
t_{low}\sim (a_{0}/2b_{0})\leq 1$, when the peaks have maximum amplitude
(for $U\lambda ^{2}/A\sim 2$). Two examples are shown in Fig. 2.

We observe that the limit $d/\lambda \rightarrow 1_{+}$ for the ratio
between the domain width and the domain wall width, is indicative of a CDW
commensurability transition from the present low incommensurable (LI) phase
to some incommensurable one. Because the validity of this LI phase model
requires that $d>\lambda $, we get the condition 
\begin{equation}
x\leq \frac{8}{5}\frac{b_{0}t}{a_{0}-b_{0}t}\ ,
\end{equation}
which delimits region LI from above in the phase diagram displayed in
Fig. \ref{fig1}.

\begin{figure}
{\centering \resizebox*{2.5 in}{!}{\includegraphics{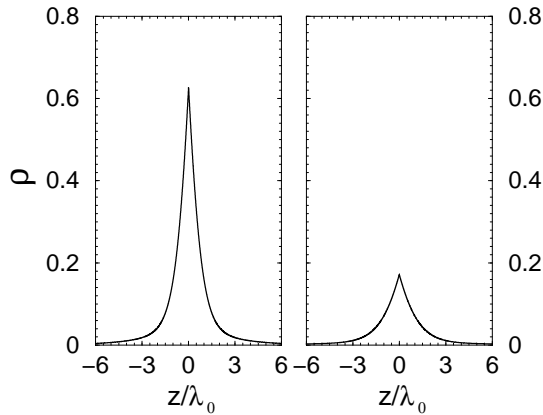}} \par}
\caption{\label{fig2}   CDW solution for $a_{0}/b_{0}=1.5$. On the left: $U\lambda
^{2}/A=1$ at $t=0.58$, with arbitrary parameters $xd/\lambda =1$ and $%
V\lambda ^{2}=0.2$. We define an arbitrary scale $\lambda _{0}$ for the peak
width at this temperature. On the right: rescaling the reduced temperature
to $t=0.31$, we have $U\lambda ^{2}/A=0.25$, $xd/\lambda =0.25$ and $%
V\lambda ^{2}=0.05$, with $\lambda =1.69\,\lambda _{0}$.   }
\end{figure}

\section{Stripe phase transition}

In the spirit of GL commensurability transitions, we now pursue the
phenomenological description of the stripe critical regime, where magnetic
order is absent. For high incommensurability, the appropriated OP is $%
\delta \rho \,e^{i\phi }$, which describes the charge modulation around the
uniform high-symmetry phase charge distribution $\rho _{0}$. In the vicinity
of the critical point $T_{S}$, the modulation is soft and therefore we
assume that $(\dot{\delta \rho })=0$, and we describe the CDW as a periodic
modulation with constant amplitude $\rho (z)-\rho _{0}=\delta \rho
\,e^{i\phi (z)}$.

The generalized GL free energy density is written as a contribution of an
explicitly dependent term on the OP derivatives $K$ plus a potential $U({\bf %
\varphi })$. The most general ``kinetic'' term has the form 
\begin{equation}
K\equiv \sum_{ij}[b_{ij}(\varphi _{i}\dot{\varphi _{j}}-\varphi _{j}\dot{%
\varphi _{i}})+a_{ij}\dot{\varphi _{i}}\dot{\varphi _{j}}]\ .
\label{kinetic}
\end{equation}
The invariant part of the antisymmetric term in (\ref{kinetic}) is called
the Lifshitz invariant, and is the essential ingredient to stabilize a
modulated incommensurate OP phase\cite{to}. The Lifshitz invariant comes
from the most general expansion which includes first order derivatives,
after one removes contributions which are not extensive in character (terms
which contribute to the total free energy $F$ as surface integrals).

The point group of the OP is chosen with a low symmetry, just noting that a
sinusoidal charge distribution locally displays mirror symmetry in the phase 
$\phi $ at maxima and minima of the distribution. Thus, with an appropriate
phase choice, we include in the symmetry group a reflection $\sigma $ to
describe the operation $\phi \rightarrow -\phi $ and the point group is $%
{\bf C}_{1v}=\{e,\sigma \}$, where $e$ is the identity\cite{hamermesh}. The
real components of the OP, $(\delta \rho \cos \phi ,\delta \rho \sin \phi )$%
, transform differently: $\delta \rho \cos \phi $ is invariant and $\delta
\rho \sin \phi $ transform after the antisymmetric representation.
Therefore, the polynomial basis to generate invariants of the group is $%
\{(\delta \rho )^{2},\delta \rho \,\cos \phi \}$. We observed that the
latter basis displays particle-hole symmetry, since 
\[
-\delta \rho \,\cos \phi =\delta \rho \,\cos \left( \phi -\pi \right)
=\delta \rho \,\cos \phi ^{\prime }\quad , 
\]
{\em i.e.} the change in sign of the charge may be included in a
redefinition of the phase variable (interchanges maxima and minima of the
charge distribution, leaving the free energy invariant).

If we write the `kinetic' energy (\ref{kinetic}) using the real components
of the OP, and using the fact that $(\dot{\delta \rho })=0$, we get $K$ in
its most symmetric form, with functions in front of the derivatives which
are invariants of the symmetry group. Note that $(\delta \rho )^{2}\cos
^{2}\phi $ is not included, if we additionally require the `kinetic' term to
be invariant under an arbitrary phase shift: 
\begin{equation}
K\equiv -a_{01}(\delta \rho )^{2}\dot{\phi}+a_{0}(\delta \rho )^{2}(\dot{\phi%
})^{2}\ .  \label{elastic}
\end{equation}
The above `kinetic' Hamiltonian (\ref{elastic}) represents the elastic
energy due to the distortion of the OP in relation to the homogeneous case.
The phenomenological parameter $a_{01}$ is related to the Lifshitz
invariant, which is linear in $\dot{\phi}$. It is then clear the role of
this term: while `kinetic' and `potential' energies compete, the Lifshitz
invariant will try to stabilize a phase with a non vanishing $\dot{\phi}$.

The `potential energy' is expanded as a power series of the polynomial basis
of invariants $\{(\delta \rho )^{2},\delta \rho \,\cos \phi \}$. This way,
the most general form of the potential up to third order is 
\begin{eqnarray}
U &\equiv &a_{1}\delta \rho \,\cos \phi +a_{2}(\delta \rho
)^{2}\,+a_{21}(\delta \rho )^{2}\,\cos ^{2}\phi \,+  \nonumber \\
&&\,a_{3}(\delta \rho )^{3}\,\cos \phi +\,a_{31}(\delta \rho )^{3}\,\cos
^{3}\phi +a_{4}(\delta \rho )^{4}\quad .  \label{potential}
\end{eqnarray}
In first place, note that particle-hole symmetry (which has been used in the
choice of the base), excludes a term of the type $(\delta \rho )^{3}\,\cos
^{2}\phi $. This is a key point in relation to the order of the transition
at $T_{S}$. More on this later. As usual in GL theories, the second order
coefficients $a_{2}$ and $a_{21}$ are assumed to be strongly temperature
dependent. At last, concerning the fourth order terms in (\ref{potential}), a comment
is pertinent. Our aim in this section is to describe the transition between
the harmonic incommensurate phase (HI in Fig. 1) and the high symmetry phase
(HS in Fig. 1), where the OP amplitude $\delta \rho $ is small (note that
the Lifshitz invariant term is proportional to $(\delta \rho )^{2}$). A
single fourth order term is then included to provide for a global stability
of our solution, assuming $a_{4}>0$. Note that a different physics may be
incorporated with other fourth order terms. Take for instance the case of $%
(\delta \rho )^{4}\,\cos ^{4}\phi $, with a negative coefficient. Clearly,
this is a `lock-in' term that should favor a homogeneous commensurate
solution for large $\delta \rho .$ For simplicity, this case will not be
considered here.

The free energy $\left( K+U\right) $ does not explicitly depend on the
variable $z$. In this case, the variational Euler-Lagrange minimization
leads to a first integral of the form 
\begin{equation}
\Lambda =\left( K+U\right) -\dot{\phi}\frac{\partial \left( K+U\right) }{%
\partial \dot{\phi}}\ ,  \label{euler}
\end{equation}
where $\Lambda $ is a constant. The explicit calculation of (\ref{euler})
results in the equation 
\begin{eqnarray}
\left( \dot{\phi}\right) ^{2} &=&\Omega \left[ \left( 1-\frac{v}{\Omega }%
\sin ^{2}\left( \frac{\phi }{2}\right) \right) +\frac{a_{21}}{a_{0}\Omega }%
\cos ^{2}\phi \,+\right.  \nonumber \\
&&\qquad \qquad +\left. \frac{a_{31}\delta \rho }{a_{0}\Omega }\cos ^{3}\phi
\right] \quad ,  \label{ELphi}
\end{eqnarray}
which couples $\phi $ with $\delta \rho $, with the definitions 
\begin{equation}
\frac{v}{2}\equiv \frac{a_{1}+a_{3}(\delta \rho )^{2}}{a_{0}\delta \rho }%
\quad ,  \label{nu}
\end{equation}
and 
\begin{equation}
\Omega \equiv \frac{a_{1}\delta \rho +a_{2}(\delta \rho )^{2}+a_{3}(\delta
\rho )^{3}+a_{4}(\delta \rho )^{4}-\Lambda }{a_{0}(\delta \rho )^{2}}
\label{omega}
\end{equation}
which are constants that parametrize our solutions.

If we take the limit $a_{21}$, $a_{31}\rightarrow 0$ in (\ref{ELphi}), the
remaining expression is a time-independent sine-Gordon equation, which
describes highly incommensurable solutions in the limit $v/\Omega \ll 1$
(note that $v/\Omega \sim \delta \rho $). For finite but small corrections
in $\left( a_{21},a_{31}\right) $, we get 
\begin{eqnarray}
u &=&\int_{0}^{\phi /2}\!{\rm d}\varphi \,\,\frac{1}{\sqrt{1-\frac{v}{\Omega 
}\sin ^{2}\varphi }}\,=\,\frac{1}{2}\Omega ^{\frac{1}{2}}z\,+  \nonumber \\
&&\qquad  \frac{a_{21}}{8a_{0}\Omega }\left( 1+\frac{3}{4%
}\frac{v}{\Omega }\right) \left( \phi +\frac{1}{2}\sin (2\phi )\right)  
\nonumber \\
&&+\frac{1}{4a_{0}\Omega }\left( a_{31}\delta \rho -\frac{3}{16}%
a_{21}\frac{v}{\Omega }\right) \sin \phi \left( 1-\frac{1}{3}\sin ^{2}\phi
\right) \,.\nonumber\\
&&   \label{ELphi2}
\end{eqnarray}
The left hand side of (\ref{ELphi2}) is an elliptical integral of the first
kind \cite{mor}. For $a_{21}$, $a_{31}\rightarrow 0$, $\phi (z)=2\,am\left(
u\right) \sim \Omega ^{1/2}z\equiv k_{0}\,z,$ for $\frac{v}{\Omega }\ll 1$,
where $\,am$ is the amplitude of the Jacobian Elliptic function. Carrying on
this procedure perturbatively up to terms of the order of $(\delta \rho )^{3}
$, we get 
\begin{eqnarray}
&\phi (z)&\,\sim \,kz+\frac{a_{21}}{8a_{0}\Omega }\left( 1+\frac{3}{4}\frac{v%
}{\Omega }\right) \,\sin (2k_{0}z)\,+  \nonumber \\
&&\,\,\frac{1}{2a_{0}\Omega }\left( a_{31}\delta \rho -\frac{3}{16}a_{21}%
\frac{v}{\Omega }\right) \sin (k_{0}z)\times \nonumber\\
&& \qquad\qquad\qquad \times \left( 1-\frac{1}{3}\sin
^{2}(k_{0}z)\right)   \nonumber \\
&&\,\equiv \,kz+\epsilon (z)\quad ,  \label{Epsilon}
\end{eqnarray}
with 
\[
k\approx k_{0}\left[ 1+\frac{a_{21}}{4a_{0}\Omega }\left( 1+\frac{3}{4}\frac{%
v}{\Omega }\right) \right] \quad .
\]
From this we can compute the CDW energy $E=l\int_{0}^{l_{z}}{\rm d}z\,(K+U)$
and obtain $k$ and $\delta \rho $ which stabilize the stripe phase at the
global minimum condition $\partial E/\partial \left( \delta \rho \right)
=\partial E/\partial k=0$. Since $\Omega \sim 1/\delta \rho ^{2}$, the terms 
$a_{21}/\Omega $ and $a_{31}\delta \rho /\Omega $ are of orders 2 and 3 in $%
\delta \rho $, respectively. This implies that the only term that carries
corrections from $\epsilon $ of the order of $(\delta \rho )^{3}$ in the
free energy, is $a_{1}\delta \rho \ \cos \left( kz+\epsilon \right) ,$ which
expanded within this order becomes 
\begin{eqnarray}
&&a_{1}\delta \rho \,[\cos (kz)-\sin (kz)\,\epsilon ]\,=\,a_{1}\delta \rho
\,\cos (k_{0}z)+  \nonumber \\
&&\qquad  -\,a_{1}\left( \frac{a_{21}\delta \rho }{8a_{0}\Omega }%
\right) \sin (k_{0}z)\sin (2k_{0}z)+{\it o}(\delta \rho ^{4})\ .
\end{eqnarray}
For $l_{z}\gg 2\pi /k_{0}$, the integrals $\int_{0}^{l_{z}}\!{\rm d}%
z\,\{\sin (k_{0}z)\sin (2k_{0}z)\}$ and $\int_{0}^{l_{z}}\!{\rm d}z\,\cos
^{n}(kz)$, with $n$ odd, are non-extensive quantities in the energy and may
be ignored. The energy will be given by the simple expression 
\begin{equation}
E=ll_{z}\left[ (\delta \rho )^{2}\left( -a_{01}k+a_{0}k^{2}+a_{2}+\frac{%
a_{21}}{2}\right) +\bar{a}_{4}(\delta \rho )^{4}\right] \ ,  \label{energy}
\end{equation}
where $\bar{a}_{4}$ is a renormalized fourth order coefficient, which
includes small contributions from $a_{31}$ and $a_{21}$. Therefore, we
assume that the condition $\bar{a}_{4}>0$ is fulfilled. Note that third
order terms in $\delta \rho $ do not contribute to $E$ in the thermodynamic
limit. The expression (\ref{energy}) is a minimum for 
\begin{equation}
k=\frac{a_{01}}{2a_{0}}\ ,
\end{equation}
and 
\begin{equation}
\delta \rho =\left( \frac{a_{01}^{2}-2a_{0}(a_{21}+2a_{2})}{8a_{0}\bar{a}_{4}%
}\right) ^{\frac{1}{2}}\ .  \label{deltarho}
\end{equation}

In the limit of highly incommensurable solutions, the potential $U$ is
irrelevant for the stabilization of the ground state. The stability is
reached through the competition of the $a_{0}$ squared gradient term (which
usually fixes the scale of energy in such classes of phenomenological
models), with the Lifshitz invariant coefficient $a_{01}$, what we
interpret as a minimization of the CDW elastic energy.

When $T\rightarrow T_{S}$ from below, $\delta \rho \rightarrow 0$, and the
Lifshitz invariant coefficient $a_{01}$ competes with the second order
coefficients $\left( a_{2},a_{21}\right) $ of the potential $U$, meaning they `cooperate' 
with $a_0$. But far
below the stripe critical point $T_{S}$, the importance of the Lifshitz
invariant term is small in comparison to the `potential' energy contribution
of the several cosine terms, which are known to stabilize LI solutions, as
we have shown in Section II. Therefore, the second order coefficients $%
\left( a_{2},a_{21}\right) $ should change sign in between of these two
limits. Thus, the precedent analysis requires the coefficients $a_{2}$ and $%
a_{21}$ to be defined in terms of a new critical parameter $\tau \equiv
T/T_{0}-1$, $T_{0}<T_{S}$, which drives a CDW commensurability transition.

Writing $a_{2}=a\tau \,$ and $a_{21}=a^{\prime }\,\tau $ ($a,\ a^{\prime }>0$%
), the stripe phase transition at $T_{S}$ will be of second order if $T_{0}$
and $T_{S}$ are related by the constraint 
\begin{equation}
T_{S}=T_{0}\left( 1+\frac{a_{01}^{2}}{2a_{0}(a^{\prime }+2a)}\right) >T_{0}\
.  \label{constraint3}
\end{equation}

Replacing (\ref{constraint3}) in (\ref{deltarho}) we find

\begin{equation}
\delta \rho =\sqrt{\frac{2a_{0}(a^{\prime }+2a)+a_{01}^{2}}{8a_{0}\bar{a}_{4}%
}}\left( \frac{T_{S}-T}{T_{S}}\right) ^{\frac{1}{2}}\ ,
\end{equation}
which goes to zero at $T_{S}$ under the usual mean field square-root law.

The periodic $\epsilon (z)$ function in (\ref{Epsilon}) introduces
non-periodical corrections in the CDW phase $\phi =kz+\epsilon (z)$. The
only way to preserve the CDW periodic nature is to set $\epsilon (z)\equiv 0$%
, {\it i.e} $a_{21}$, $a_{31}=0$. In this picture, the CDW solution is just $%
\rho (z)=x+\delta \rho \cos (%
{\displaystyle {a_{01} \over 2a_{0}}}%
\,z)$, using appropriate units, with $x$ being the doping. Applying the
global constraint $l\int_{0}^{l_{z}}\rho (z)\,{\rm d}z=Q$ in the limit $%
l_{z}\gg \frac{2\pi }{k}$, we encounter that $\rho _{0}=x$. Since $\delta
\rho \leq \rho _{0}$, the $(\dot{\delta \rho })\approx 0$ region is
delimited by the inequality $x\geq \delta \rho $ in the phase diagram
depicted qualitatively in Fig. 1.

\section{Discussion}

We present here a brief summary of the paper and we discuss the results of
the last two sections. Phase I describes the LI spin driven regime of
stripes, which takes place in our phase diagram for low doping
concentrations. This phase is also experimentally associated with
incompressible stripes (for $x<\frac{1}{8}$, in the cuprate case). This is
an evidence that they interact weakly, despite the existence of `long-range'
spin-spin and charge-charge interactions (in both cases, the correlation
length is of the order of few hundred angstroms). Separately, each of them
would drive the system to macroscopically homogeneous phases. However, their
competition, even in the absence of dynamical correlations, reduces
considerably the effective range of the interactions and stabilizes
non-homogeneous phases in a smaller scale.

We showed in Section I that a simple static Hamiltonian, like the one given
by (\ref{hamiltonian1}), captures in a qualitative way the physics of
competing effects which leads to low temperature CDW formation in the
presence of two-dimensional magnetism. The second conclusion is that this
soliton-like CDW scenario changes drastically with temperature. Moving in
region LI at constant doping and increasing temperatures (see Fig. 1), our
solution shows that the ratio between the inter-stripe distance and the
stripe width decreases (as well as the amplitude of the peaks) and goes to
the limit $d/\lambda \rightarrow 1$, which corresponds to the
commensurability transition that separates the LI phase from the high
incommensurable (but still spin driven) phase I.

In contrast, in the absence of 2D antiferromagnetism, we may ask which is
the mechanism behind the CDWs, since there are no magnetic correlations to
compete with the repulsion of holes. For this phase, we construct the free
energy from general theoretical arguments coming entirely from symmetry. We
conjecture that this free energy includes, in a stationary fashion,
dynamical phenomena that occur at high temperature, such as elastic and
electronic coupled effects that induce a polarization of the lattice. In an
actual phase diagram of cuprates, as the one shown in Fig. 4 of Ref.%
\onlinecite{torrance}, this phenomenon is located in the high temperature
sector, considerably far from the tetragonal-orthorhombic structural phase
transition. This latter is still in the interior of the 2D
antiferromagnetic correlated region, and has no relation with the high
temperature physics that we are dealing with here.

The $C_{1v}$ point group of the OP for the HI phase implies the occurrence of
odd terms in the free energy density. However, particle-hole symmetry makes
those terms non-extensive in character, and they can be neglected in the
thermodynamic limit. This means, in principle, that one may not rule out the
possibility of a second order phase transition for stripes, in contrast to
other phenomenological approaches\cite{za}.

\acknowledgments
The authors are grateful to {\em Funda\c{c}\~{a}o de Amparo \`{a} Pesquisa
do Estado de S\~{a}o Paulo }(FAPESP, Brazil) for partial financial support
through the project 00/06881-9. One of the authors (GGC) would also like to
acknowledge support from FAEP (Unicamp, Brazil).

\section*{Appendix}

In this appendix we will demonstrate that the GL free energy (\ref
{freeEnergy2}) is equivalent to Hamiltonian (\ref{hamiltonian1}) in mean
field approximation. Starting from the definition of our staggered mean
field ${\bf m}$, we calculate macroscopic average of the spin term in (\ref
{hamiltonian1}), which yields 
\begin{eqnarray}
{\rm H}_{s} &=&-\int \!{\rm d}^{2}r\,\,\,J\sum_{j=1}^{2}\,\alpha
_{j}m_{j}^{2}({\bf r})  \nonumber \\
&-&\int \!\!\!\int \!{\rm d}^{2}r\,{\rm d}^{2}\epsilon \,\,\,J^{\prime
}\sum_{j=1}^{2}\,\beta _{j}m_{j}({\bf r})m_{j}({\bf r}+{\bf \epsilon })\ ,
\end{eqnarray}
in the limit where the lattice parameter goes to zero. The charge term
calculation is similar, and we may write 
\[
{\rm H}_{c}=\int \!\!\!\int \!{\rm d}^{2}r\,{\rm d}^{2}\epsilon \,\,\frac{1}{%
\epsilon }\left[ 1-\eta ({\bf r})\right] \left[ 1-\eta ({\bf r}+{\bf %
\epsilon })\right] 
\]
Observing that the OP is by definition a continuous function that varies
slowly in space, we assume that $\eta ({\bf r}+{\bf \epsilon })$ and ${\bf m}%
({\bf r}+{\bf \epsilon })$ can be expanded up to terms of lowest order.
Performing the integration in $\epsilon $, the resulting free energy is a
functional of the form 
\begin{equation}
f({\bf r})=-a_{o}(\dot{\eta})^{2}+a_{2}(1-\eta )^{2}+\sum_{j=1}^{2}\left[
b_{o}(\dot{m}_{j})^{2}-b_{2}\alpha _{j}m_{j}^{2}\right] \ ,
\label{freeEnergy}
\end{equation}
with $\alpha _{j}$ taken as an effective anisotropy parameter.

Note that the charge $(1-\eta )$ acts as a secondary order parameter, with
no critical behavior. In order to achieve a minimum for the thermodynamic
potential, we must include higher order powers of the primary OP ${\bf m}$
in the above expression. Since the odd ones break the free energy invariance
under reflections, the lowest correction is of fourth order. So, we
heuristically introduce the simplest term $b_{4}\,m^{4}$, which represents
an isotropic magnetic quadrupole interaction. In the same way, we may
consider a coupling term for the spin and charge degrees of freedom. We rule
out the OP coupling with odd powers of $(1-\eta )$, due to particle-hole
symmetry. If we assume that the coupling does not depend on the crystal
directions, the lowest order term is $c_{2}\,(1-\eta )^{2}\,m^{2}$. This
closes expression (\ref{freeEnergy2}) in our model.

\end{document}